# Vibrational Modes and Terahertz Physical Phenomena Underpinning ZIF-71 Metal-Organic Framework

*Annika F. Möslein [1] and Jin-Chong Tan [1]**

[1] Multifunctional Materials and Composites (MMC) Laboratory, Department of Engineering Science, University of Oxford, Parks Road, Oxford OX1 3PJ, United Kingdom

*Corresponding author: jin-chong.tan@eng.ox.ac.uk

**ABSTRACT**

The zeolitic imidazole framework ZIF-71 has the potential to outperform other well-studied metal-organic frameworks due to its intrinsic hydrophobicity and large pore size. However, a detailed description of its complex physical phenomena and structural dynamics has been lacking thus far. Herein, we elucidated all vibrational modes of ZIF-71 using high-resolution inelastic neutron scattering and synchrotron radiation infrared spectroscopy in conjunction with density functional theory calculations. We discover low-energy collective modes, such as gate-opening and shearing mechanisms that may affect the functions and framework stability of ZIF-71. Its single-crystal mechanical properties are further unraveled by nanoscale analytics.

**KEYWORDS:** Metal-organic frameworks, lattice dynamics, density functional theory, infrared spectroscopy, inelastic neutron scattering, nanomechanical behavior

Amongst the vast field of nanomaterials, metal-organic frameworks (MOFs) have gained considerable interest owing to their unique physical and chemical properties, which are unattainable in other conventional materials. For instance, their open framework structure assembled from metal clusters bridged by organic linkers leads to large surface areas even exceeding those of zeolites, while their organic-inorganic character offers novel, tailorable functional properties.[1] Originating from the traditional use of porous nanomaterials, where MOFs have been proven beneficial for gas capture and storage, the multifunctional nature of MOFs has paved the way for an array of innovative applications, including but not limited to catalysis, drug delivery, microelectronics, and chemical sensors.[2-7]

One of the most promising candidates for the application of MOFs is the zeolitic imidazole framework, ZIF-8 [$Zn(mIM)_2$; mIM = 2-methylimidazolate], due to its stability and ease of synthesis.[8] ZIF-8 crystallizes in a sodalite (SOD) topology with an internal pore size of ~10 Å, and it has become a prototypical and well-studied material among the large family of MOFs.[9] ZIF-8, or materials in the subclass of ZIFs, in general, are constructed from metal cations tetrahedrally coordinated to imidazole-type organic linkers, yielding a chemically stable framework structure with cage-like subunits.[8] While ZIF-8 has indeed sparked considerable scientific and technological interests, other ZIF materials, in fact, might even outperform ZIF-8 in various applications.[10,11] For instance, the far less studied material ZIF-71, built from Zn cations bridged by 4,5-dichloroimidazolate (dcIM) linkers, encompasses the RHO-type cages with pore sizes exceeding those of the SOD-type ZIF-8, thus rendering ZIF-71 a promising candidate for enhanced gas capture or mechanical shock absorbance.[12] It crystallizes in a cubic symmetry, and is constructed from large $\alpha$-cages (16.5–16.8 Å of diameter) connected by eight-membered ring (8MR) units with cage windows of 4.2–4.8 Å, in addition to four- and six-membered ring (4MR, 6MR) pore apertures (see Fig. 1). In addition, the co-existence of hydrogen and chlorine atoms in the dcIM linker offers more versatile interactions with guest molecules than expected for only hydrogen bonds in ZIF-8, advancing reactivity in catalysis, or selectivity for sensing applications. Yet, perhaps owning to its complex structure, this material has not been widely explored which is surprising given that ZIF-71, due to its intrinsic hydrophobicity, provides the excellent chemical stability akin to ZIF-8. While a thorough understanding of the physical properties of ZIF-8 and its underpinning lattice dynamics has been developed,[13,14] little is known about the fundamental vibrational characteristics of ZIF-71, which is so central to understanding the physical behavior of the material, and thus there is a gap in knowledge prior to targeting specific applications.

In this work, we provide the first complete assignment of the vibrational modes of ZIF-71 using high-resolution neutron and synchrotron vibrational spectroscopy, in conjunction with *ab initio* quantum mechanical simulations. This multimodal approach allows us to establish the low-frequency terahertz (THz) lattice modes and unravel basic mechanistic dynamics, and further identifies the characteristic vibrational modes in the mid-infrared region. The results are important because not only they provide the missing reference for spectroscopic studies of ZIF-71, but also they hold the key to unlocking complex host-guest interactions underpinning the functions of ZIF-71.

To analyze the physical molecular vibrations corresponding to each mode of ZIF-71, we computed the theoretical spectra and vibrational frequencies using density functional theory (DFT), as implemented in a development version of the CRYSTAL17 code (running in massively parallel processing (MPP) mode on high-performance clusters).[15,16] To the best of our knowledge, ZIF-71, with its 816 atoms per unit cell, is hitherto the largest MOF system for which theoretical frequency calculations with DFT have been accomplished. We have tested two all-electron basis sets of increasing size, designated as BS1, and BS2, containing 12,480, and 16,032 local functions, respectively. The calculations were performed at the B3LYP-D3 level of theory, including two- and three-body corrections (ABC) to account for dispersion interactions.[17-19] The Fourier transform infrared spectroscopy (FTIR) data were obtained with synchrotron radiation (SR) at the MIRIAM beamline at the Diamond Light Source (Oxfordshire, UK). Using two different detectors (bolometer and built-in detector) and beam splitters (Mylar and KBr, respectively), the full broadband IR spectrum from 50 – 2000 $cm^{-1}$ could be measured. For IR spectroscopy, the interaction between electromagnetic waves and molecular vibrations is based on dipole changes, which are only induced by asymmetric vibrations or rotations leading to the so-called selection rule. To further elucidate the symmetric modes without dipole change, or the IR inactive modes, these datasets were complemented with inelastic neutron scattering (INS) measurements, performed on the TOSCA spectrometer at ISIS Neutron & Muon Spallation Source (Oxfordshire, UK).[20] Unlike optical spectroscopy techniques, all molecular motions are observed in INS without the symmetry-based selection rule, however, in practice, this technique shows dominant sensitivity to vibrations encompassing hydrogen due the exceptionally large scattering cross-section of the hydrogen nuclei.[21] Additionally, the vibrational dynamics in the low-energy THz region, which are so central to the structural mechanics of MOF materials, are revealed with INS, as frequencies as low as 20 $cm^{-1}$ are measured. We further employed nanoscale analytics, such as infrared nanospectroscopy and nanoindentation, to attain the local chemical and physical information of the individual ZIF-71 crystals. Both techniques are based on atomic force microscopy (AFM), albeit operated in different modes: nanoindentation monitors the strain rates of the AFM indenter tip during the indentation process to probe the local mechanical properties, specifically the Young's modulus ($E$) and hardness ($H$) of single crystals.[22] Nanospectroscopy is based on a tapping-mode AFM combined with a scattering-type scanning near-field optical microscope, where the illuminated tip serves as a source for an evanescent near-field, to obtain a nanoFTIR spectrum of individual nanocrystals.[23-25] Together, these multimodal techniques gave us a detailed "picture" of ZIF-71, comprising its intrinsic vibrational dynamics, its fundamental physico-chemical behavior, and the resulting single-crystal characteristics.

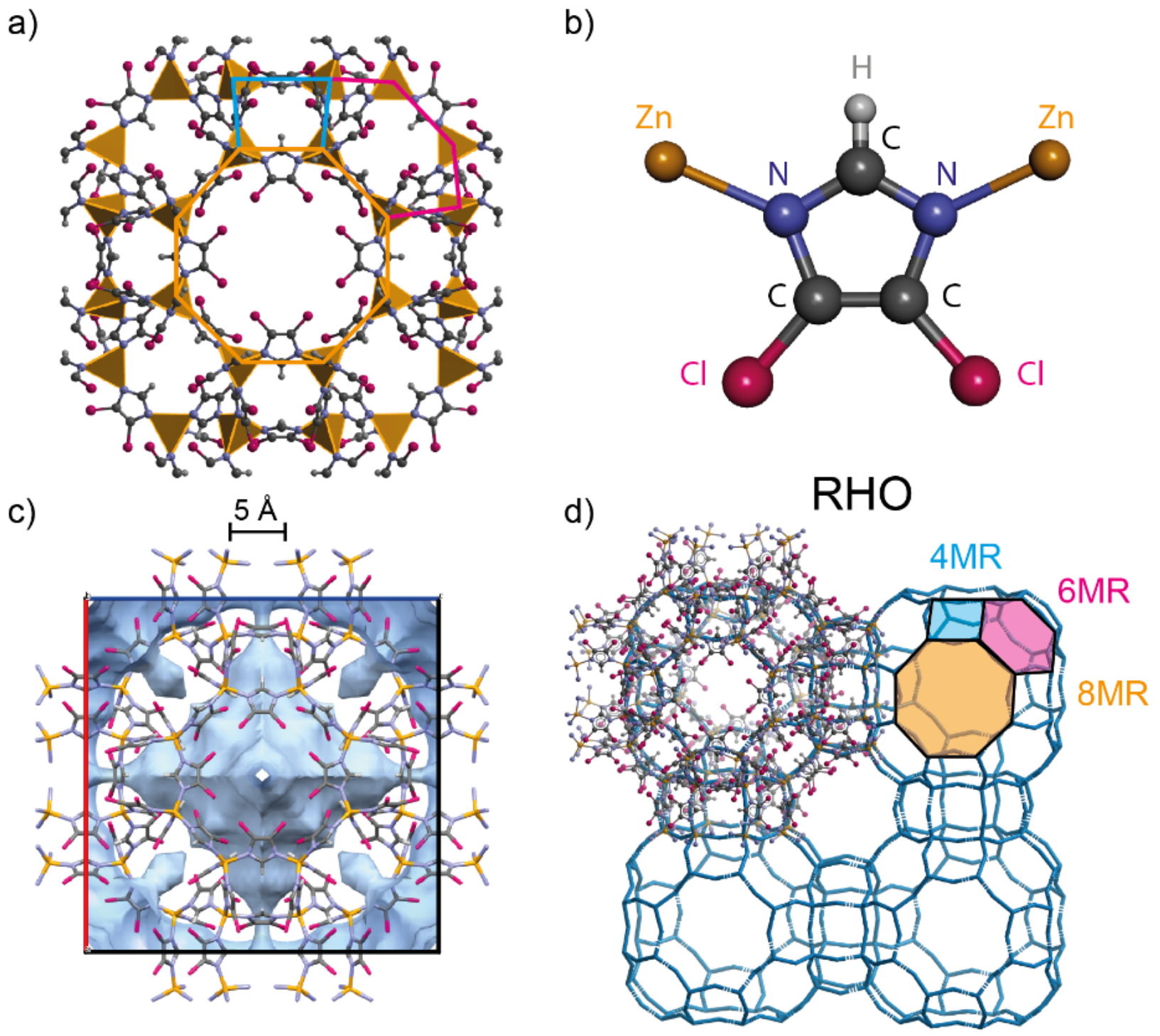

FIG 1: Framework structures of ZIF-71: a) ZIF-71 unit cell where the inorganic building blocks are illustrated by the $ZnN_4$ tetrahedra. b) The building unit showing the Zn–dcIM–Zn linkages. c) The blue surfaces denote the nanopore, corresponding to the solvent accessible volume within the open framework structure. d) Illustration of the RHO topology, highlighting the apertures of the 4-, 6-, and 8-membered rings (MR).

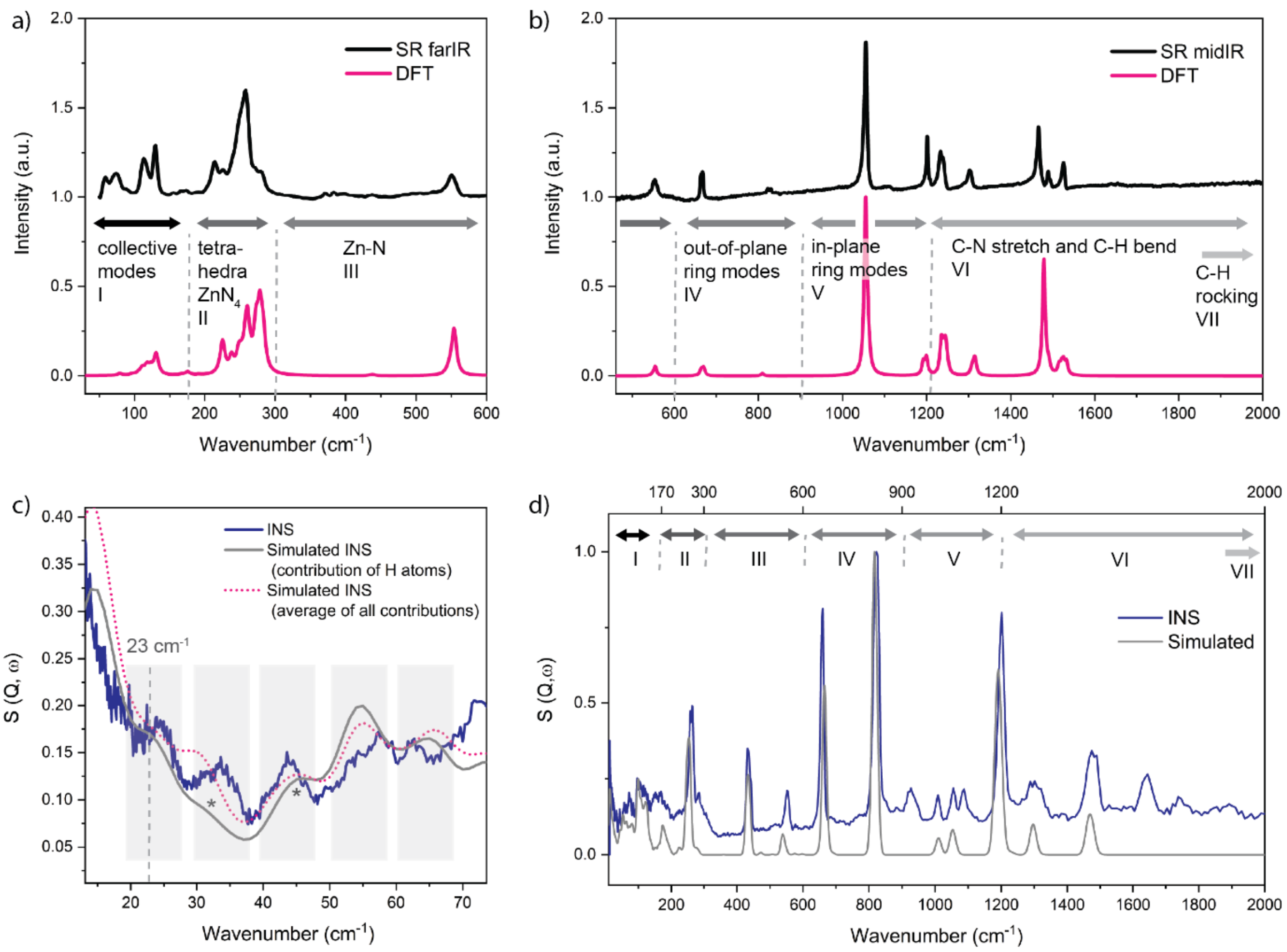

FIG 2. Comparison of experimental and theoretical DFT spectra for ZIF-71. a) Far-infrared (farIR) spectrum measured with synchrotron radiation (SR) compared with DFT simulated spectrum. b) Mid-infrared (midIR) spectrum measured with synchrotron radiation (SR) compared with DFT simulated spectrum (shifted with a factor of 0.98). c) Low-energy region of the spectrum obtained with inelastic neutron scattering (INS), compared with calculated INS spectra derived from the DFT phonon calculation). d) Simulated and experimental INS spectra of ZIF-71.

As shown in Fig. 2, the calculated IR spectrum yields excellent agreement with the one measured with SR-FTIR. A bulk shift of the simulated peaks to lower frequencies was applied (factor 0.98), which is a common approach considering the “nanocrystal effect”, as the strengths of real bonds, even if only slightly, are decreased from the ones of idealized crystal.[26] For reference, the measured INS spectrum for ZIF-71 is also shown in Fig. 2d. It can be seen that the shape of the predicted spectrum in the low-energy region matches remarkably well with the INS data (Fig. 2c); this is a significant result given that establishing a precise agreement between DFT and INS data at low wavenumbers is usually considered as a challenge even for a less complex framework.[13,27,28] Only the combination with DFT can assign the physical motions to each observed peak, and a detailed analysis of all vibrational modes identified different characteristic spectral regions ranging from high to low energies: above 2000 $cm^{-1}$, a region typically associated with the stretching vibration of functional groups, C-H stretching modes are observed for ZIF-71. In the transition between functional group and the fingerprint region between 1200 and 2000 $cm^{-1}$, the high-intensity peaks are assigned to C-N stretching modes of the aromatic ring in combination with C-H bending. Below that, the characteristic modes of the aromatic ring of dcIM are prevalent in the mid-IR fingerprint region: 900-1200 $cm^{-1}$ for vibrations describing the in-plane ring modes, and 600-900 $cm^{-1}$ for the out-of-plane ring modes, respectively. It is further evident that the modes involving the $ZnN_4$ metal clusters appear below 500 $cm^{-1}$, where stretching and bending between Zn and N are excited at specific frequencies; here, however, the Zn atoms remain fixed and the main resulting motions are associated with the linker units, thus slightly deforming the pores and channels of ZIF-71. Stronger structural distortions of the pores, and the framework itself, are expected in the low-energy, or THz region (<300 $cm^{-1}$), involving the low-energy collective modes. This is precisely where relations between the physical properties and the lattice dynamics can be instigated, defining phenomena like gate-opening, shearing, and phase transitions. Between 170 and 280 $cm^{-1}$, the N–Zn–N bending and stretching modes are observed introducing tetrahedral deformations, which, in turn, cause distortion of the linker unit, and some – albeit small – structural deformations of the pores will occur. Here, vibrations associated with the Cl atoms are also detected; they can play a key role for they offer additional interaction sites for guest adsorption. Stronger deformations of the 4-, 6-, and 8-membered rings (MR) are revealed in the spectral region below 150 $cm^{-1}$ ($\lesssim$ 4.5 THz): this is where intriguing physical phenomena like gate-opening, shearing, pore breathing, and other structural mechanisms underpinning the fundamental properties of the framework are prevalent.

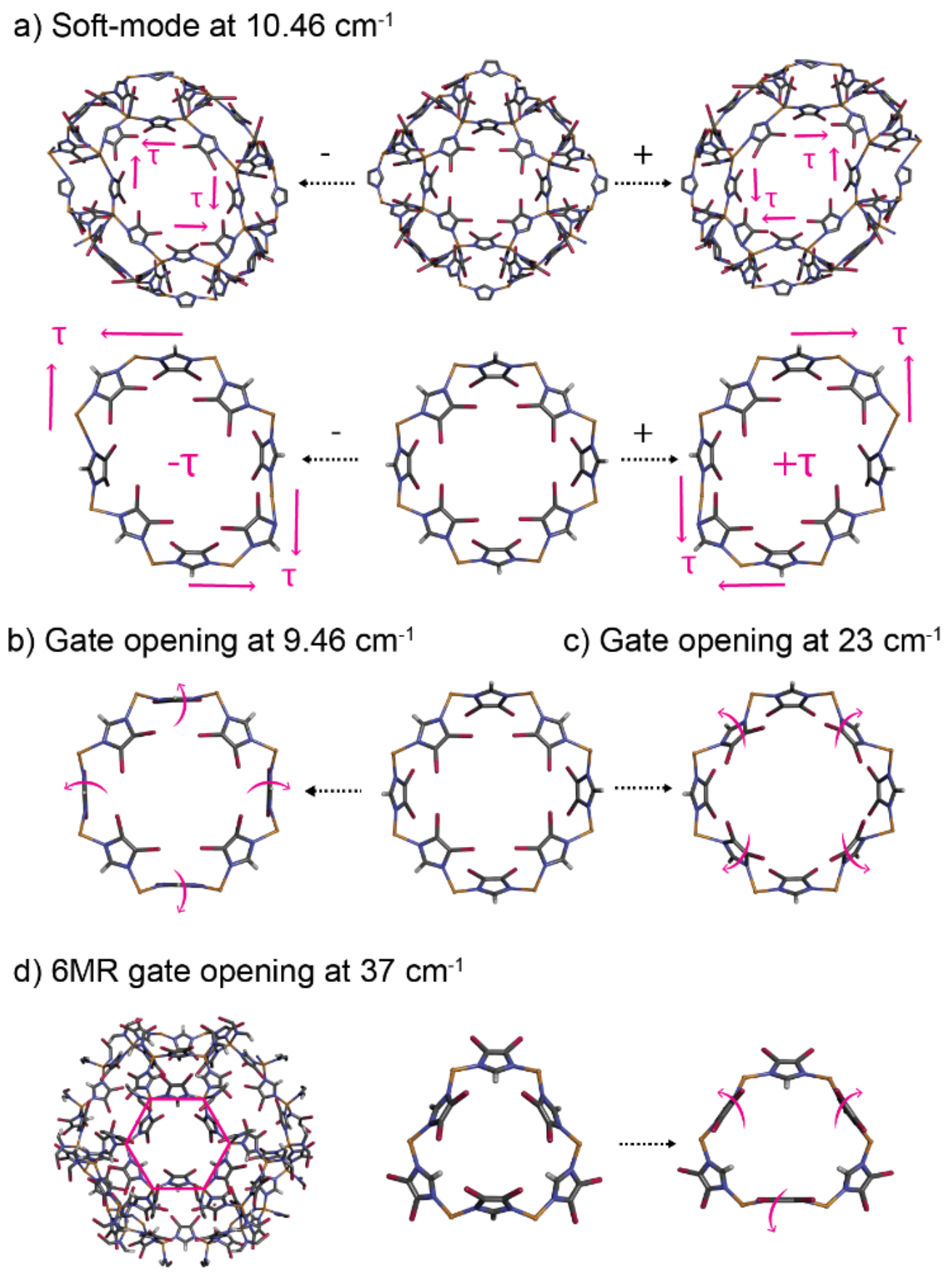

FIG 3: Low-energy lattice modes of ZIF-71. a) Soft mode caused by a shear deformation of the 8MR. b, c) Gate-opening mechanisms of the 8-membered ring (MR) *via* synchronous rocking of opposite organic linkers. Pink arrows designate the collective dynamics affecting the geometry of pore cavity. d) Gate-opening of the 6MR through synchronous flapping of the linker units.

Herein, we explore the low-energy collective modes of ZIF-71, which encompass contributions from the entire crystalline lattice and thus, are so intrinsically linked with the core physical phenomena observed in ZIF materials. Whilst a full description of the vibrational modes is provided in the Supplementary Materials (SM), we illustrate in Fig. 3 a few crucial lattice modes that strike us as exceptional for understanding the physical phenomena of ZIF-71. Perhaps one of the most significant lattice dynamics among them is the soft mode at 10.46 $cm^{-1}$ (~0.3 THz), which is assigned to a strong shear deformation of the 6- and 8-membered rings (Fig. 3a). Such a shearing deformation suggests a propensity to undergo a phase transformation, potentially to ZIF-72 or COK-17, which contain, in essence, the same building blocks as ZIF-71 and yet, their structures are entirely different: the latter manifests in a SOD topology akin to ZIF-8 but with a distorted configuration, whereas ZIF-72 is a non-porous *lcs*-type framework lacking the exceptional porosity of ZIF-71.[29,30] Transitioning – or even structural amorphization as observed in other ZIF materials[13,31] – seems likely, especially since the shearing mode of the 8MR, which are inherently mechanically unstable subject to antiparallel shear forces given the large pore size, leads to a decrease of the pore that is even more susceptible to collapse.[32,33] This THz mode could explain the previously observed phase transition reported during intrusion-extrusion experiments, where, despite a major collapse of the ZIF-71 framework, traces of both ZIF-71 and ZIF-72 were found by X-ray diffraction.[34]

In addition, we discovered three collective modes triggering gate-opening, a mechanism which could facilitate adsorption and significantly raise the gas uptake capacity. This is due the synchronous flapping (or scissoring mode) of opposite ligands that results in a greater accessible pore volume *via* an opening of the pore aperture. For the 8MR, we detect two gate opening modes at 9.45 and 23 $cm^{-1}$, respectively, induced due to the conformational changes of different linker units as they pivot around the metal centers (Fig. 3b,c). Directly related to the increase in aperture of the 8MR, a similar, albeit less pronounced, pore breathing mechanism is propagated in the adjacent 6MR, whereas the 4MR exhibits shearing deformation. An actual gate-opening in the 6MR is however identifiable at 37 $cm^{-1}$ (~1 THz), where the coherent scissoring dynamics of the linker units located opposite to each other cause an increase of the pore aperture (Fig. 3d). Though much less explicit, the lattice vibrations involving stretching and twisting of the Zn–N bonds and ligands also trigger pore deformations of the 4, 6 or 8MR – be it asymmetric gate-opening, pore breathing, expansion, contraction, or shearing – and while all of these mechanisms indeed distort the cage structure, they are thus far not assigned to the core physical phenomena observed in ZIF-71. In general, we observe less structural flexibility for ZIF-71 in comparison with the SOD-type ZIF-8, where the mIM linkers offer a higher capability to twist than the dcIM moieties. In the case of ZIF-71, this could in fact facilitate the trapping of guest molecules, as a higher internal loading would not immediately lead to deformation of the pore aperture. Especially in combination with a larger cage diameter, these phenomena render ZIF-71 a very promising candidate for applications targeting the areas of guest encapsulation, gas adsorption and capture.

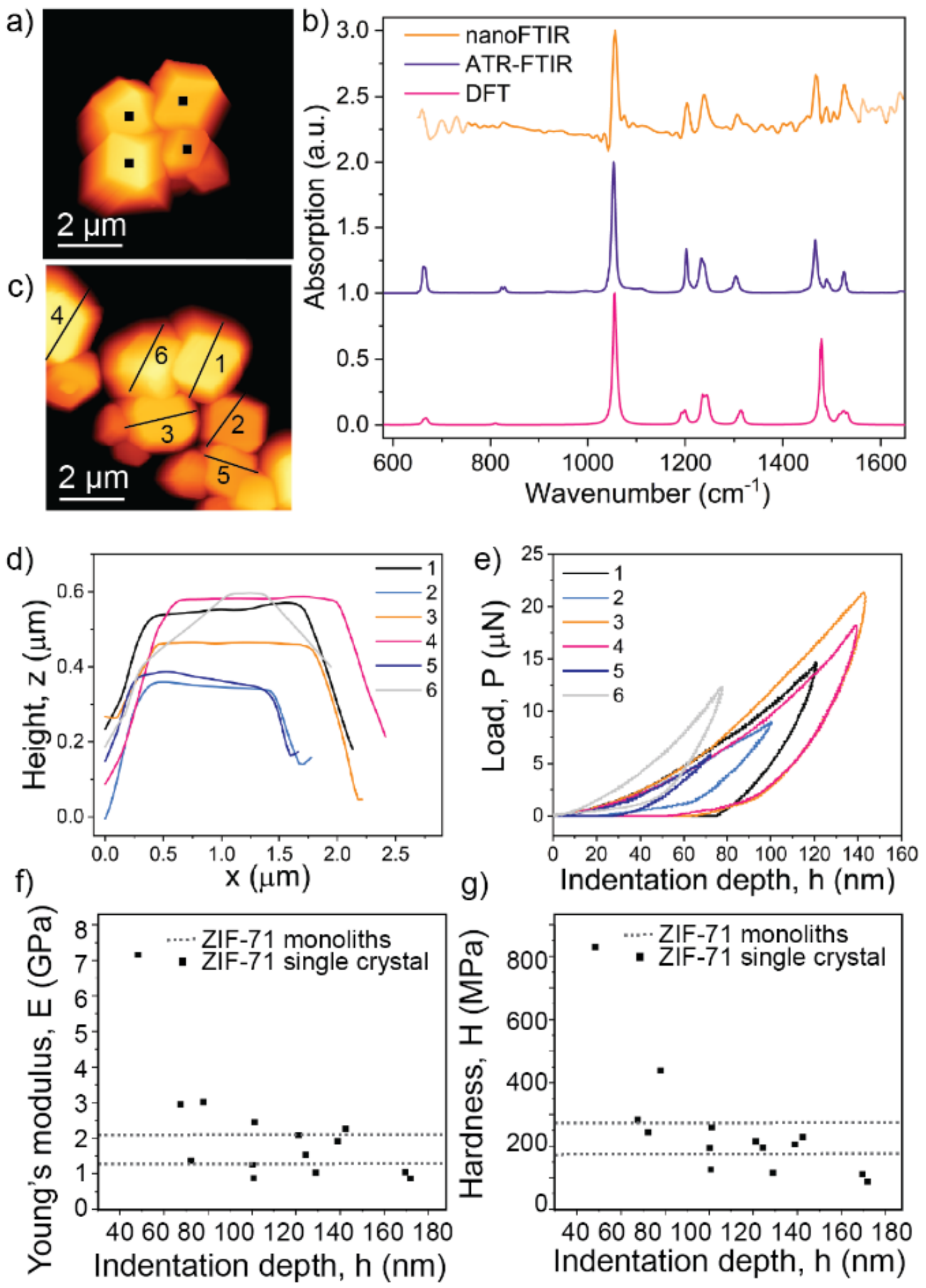

FIG 4. Nanoscale analytics of ZIF-71 single crystals. a) AFM image of ZIF-71 crystals with indicated positions for nanoFTIR measurements. b) Corresponding nanoFTIR spectra

compared with ATR-FTIR measurements and DFT simulations (shifted with a factor of 0.98). c) Individual crystals are selected for AFM nanoindentation measurements. d) AFM height profile of the individual crystals corresponding to the lines designated in c). e) AFM nanoindentation force–displacement (*P-h*) curves of individual nanocrystals of ZIF-71. f) Derived Young's modulus, and g) hardness, plotted as a function of the maximum indentation depth; dashed lines represent values measured on ZIF-71 monoliths with instrumented nanoindentation by [35].

To obtain a better understanding of the physical properties of the individual ZIF-71 crystals, we performed nanoscale analytical measurements comprising near-field infrared spectroscopy and AFM nanoindentation. First, we measure, with a resolution of 20 nm, the local IR vibrational spectra: they not only reveal the homogeneity of the chemical composition of a single crystal (see SM Fig. S5), but their average also offers comparison with conventional (far-field) ATR-FTIR techniques and the simulated spectrum, as shown in Fig. 4b. For instance, the most pronounced peak at 1054 $cm^{-1}$, assigned to in-plane ring deformation of the linker with rocking of the C–H groups, is characteristic for ZIF-71, while the smaller peaks associated with C–N stretching modes at 1201 $cm^{-1}$ (with C–H bend), 1234 $cm^{-1}$ (ring breathing), and 1301 $cm^{-1}$ (stretching) are also in agreement with conventional measurements. The small discrepancies between the theoretical and experimental spectra around 1500 $cm^{-1}$ stem from the fact that here, symmetric stretches of the under-coordinated N–C–Cl groups at the edge of the unit cell are triggered which are not expected in the ideal, periodic crystal. Thus, the peak splitting observed in the experimental data indicates that these modes are in fact present at the crystal surface. Otherwise, the local single-crystal spectrum matches the one measured on bulk, polycrystalline material and calculated from a periodic lattice; hence, this technique can be used for direct recognition of the vibrational modes, or the fingerprint, of ZIF-71, thereby facilitating prospective studies on the behavior of ZIF-71 with nanoscale resolution.

In addition, we measured the local mechanical properties of single ZIF-71 crystals by employing AFM nanoindentation. This technique allows us, with a resolution akin to AFM, to accurately characterize the elastic stiffness and hardness of individual ZIF-71 crystals, which is unachievable using standard techniques, since the growth of large single crystals of ZIF-71 crystals (~100s μm) suitable for instrumented nanoindentation has been proven challenging up to this point. Herein, we obtain a set of load-vs-displacement (*P*-*h*) curves of several isolated ZIF-71 crystals with sub-micron size. From these shallow indentations with a surface penetration depth ranging from 60 to 170 nm, the Young's modulus (*E*) lying in the range of of 1-3 GPa and hardness (*H*) between 100-300 MPa are determined using the Oliver and Pharr method, taking into account the cube-corner geometry of the diamond indenter tip.[36] It is worth mentioning that the outlier ($E$ = 8 GPa, $H$ = 800 MPa) was measured on the inclined crystal (label 6 in Fig.4 c-e), and introduces artefacts, as the lack of smooth and flat sample surface led to an unreliable contact area determination. Owing to the difficulty to grow significantly (at least 100 times) larger ZIF-71 crystals, the mechanical properties of ZIF-71 from instrumented nanoindentation or Brillouin spectroscopy have not been reported yet, and thus, it is challenging to verify these experimental values from AFM nanoindentation. However, while this is true, it is a method that has been shown to achieve quantitative measurement on the prototypical ZIF-8, where comparison with conventional techniques and DFT calculation were feasible. With the efficacy of this method being proven, we herein report the mechanical properties of ZIF-71 with nanoscale techniques, which further substantiate these findings. Additionally, the values of the Young's modulus and hardness are in reasonable agreement with the ones measured on ZIF-71 monoliths (Fig. 4f,g): more precisely, the discrepancies can be linked to various factors including nanostructure packing, anisotropic behavior of single crystals, or compliance of the AFM cantilever probe.[35] In general, the Young's modulus of ZIF-71 ($E$ ~2 GPa) is notably lower than the one previously shown for ZIF-8 (3.2 GPa), an observation which we attribute to the larger pore size of ZIF-71.

Exceptionally large cages, coupled with the versatility of their pore apertures (4, 6, and 8MR) featuring interconnected channels, is the one of the main reasons why ZIF-71 is such a promising candidate for applications, including – but not limited to – gas capture and storage, or the encapsulation of luminescent guest molecules through nanoscale confinement.[37] All of these are further facilitated by the possible interactions with the Cl atoms, which can be leveraged for targeted host-guest interactions, enabling catalytic behavior or chemical sensing. Overall, ZIF-71 offers, not only due to its hydrophobicity, the ability to outperform prototypical MOF-materials such as ZIF-8. One example is the storage or dissipation of mechanical energy, using the liquid-phase intrusion of concentrated electrolytes in a hydrophobic nanoporous framework, where the stored energy in ZIF-71 is almost doubled compared to that measured for ZIF-8.[12] Similarly, ZIF-71 performed better than ZIF-8 in more recent impact absorbance experiments based on water intrusion: the larger water network in a ZIF-71 cage is more stabilized than in the smaller ZIF-8 cage, and accordingly it is less favorable for a water molecule to hop to an empty neighboring ZIF-71 cage.[11] This phenomenon increases the intrusion barrier in ZIF-71 compared to ZIF-8. Albeit less significantly, the lower flexibility in ZIF-71 could also hinder water hopping, thereby decelerating the intrusion, and thus enhancing the mechanical energy absorption capacity of ZIF-71.

The above exemplar, however, is only one of many possible applications where ZIF-71 can outperform other, well-studied MOF materials and ZIF counterparts. For instance, ZIF-71 thin films have been shown to be promising for nanofabrication of MOF devices targeting low-*k* dielectrics and photonic sensors.[38] Our work presents the fundamental insights required prior to developing such applications and technologies by contributing a full description of the vibrational dynamics of ZIF-71. Combining DFT calculations with high-resolution synchrotron FTIR spectroscopy and inelastic neutron scattering not only completely characterizes each vibrational mode, but further unravels the key collective modes which are inherently linked with the material's properties and functions. For instance, we discovered shearing modes with potential phase transitioning and gate-opening modes of the different channels, which could increase gas uptake. In addition, we explore the single-crystal properties of ZIF-71 using nanoscale analytical tools. This allows us, while simultaneously imaging the crystals with AFM, to locally probe the chemical composition by measuring a nanoFTIR spectrum from a 20 nm spot; and we further measured the local mechanical properties to complete the detailed picture of ZIF-71. We hope to offer the basis for – and inspire – further studies on the physical behavior of ZIF-71, as a versatile platform for basic research and application stemming from its unique topology, hydrophobicity, large pore size, and nanoscale mechanics.

## ACKNOWLEDGMENTS

A.F.M. thanks the Oxford Ashton Memorial scholarship for a DPhil studentship award. J.C.T. and A.F.M. are grateful for funding through the ERC Consolidator Grant (771575 (PROMOFS)) and the EPSRC Impact Acceleration Account Award (EP/R511742/1). We thank large facilities access through the ISIS Beamtime at TOSCA (RB1910059) and the Diamond Beamtime at B22 MIRIAM (SM21472). We acknowledge the use of the University of Oxford Advanced Research Computing (ARC) facility in carrying out this work (10.5281/zenodo.22558). Via our membership of the UK's HEC Materials Chemistry Consortium (MCC), which is funded by EPSRC (EP/R029431), this work used the ARCHER2 UK National Supercomputing Service (http://www.archer2.ac.uk). We thank Lorenzo Donà for compiling the development version of CRYSTAL17 on ARCHER2, and Prof. Bartolomeo Civalleri for the provision of DFT basis sets. We are very grateful to Dr. Svemir Rudić (ISIS), Drs. Mark Frogley and Gianfelice Cinque (Diamond) for scientific discussions.

*Supplementary Materials*

*for*

# Vibrational Modes and Terahertz Physical Phenomena Underpinning ZIF-71 Metal-Organic Framework

*Annika F. Möslein[1] and Jin-Chong Tan[1]**

[1]Multifunctional Materials and Composites (MMC) Laboratory, Department of Engineering Science, University of Oxford, Parks Road, Oxford OX1 3PJ, United Kingdom

*Corresponding author: jin-chong.tan@eng.ox.ac.uk

*Table of Contents*

## 1.1 Synthesis of ZIF-71 Nanocrystals

ZIF-71 was synthesised by dissolving 2.155 g of 4,5-dichloroimidazolate (dcIm) in 114.96 ml of methanol and combining it with 0.72 g of zinc acetate, likewise, diluted in 144.96 ml of methanol. The obtained white colloidal solution was stirred for 24 hours at room temperature. Nanocrystals were isolated by centrifugation at 8000 rpm for 5 minutes and washed with fresh methanol, a procedure repeated three times to remove any excess reactants. Prior to the measurements, the material was exposed to heat (80 °C) and vacuum overnight to remove any solvent.

## 1.2 Powder X-ray Diffraction

The purity and crystallinity of the powdered sample were confirmed using powder X-ray diffraction (PXRD). XRD patterns were recorded using a Rigaku MiniFlex diffractometer with a Cu Kα source (1.541 Å). Measurements were performed with a step size of 0.02° and step speed of 0.01° $min^{-1}$.

## 1.3 FTIR Measurements with Synchrotron Radiation

Infrared (IR) absorption spectroscopy experiments were performed at the Multimode InfraRed Imaging and Microspectroscopy (MIRIAM) Beamline (B22) at the Diamon Light Source. The absorbance spectra were measured using a Bruker Vertex 80V Fourier Transform IR (FTIR) interferometer (Bruker Optics, Germany). The broadband spectral range of the synchrotron source covered the visible to the low-energy THz region. A liquid helium-cooled bolometer in combination with a 6-μm Mylar multilayer beamsplitter in the interferometer allowed for measurements in the far-IR spectral region between 0 and 650 $cm^{-1}$. Mid-IR measurements (550 – 4000 cm$^{-1}$), on the other hand, were carried out using a KBr beamsplitter and an in-built detector. IR spectra were collected with a resolution of 2 cm$^{-1}$ and 256 scans per spectral scan. All measurements were performed at vacuum (lower than 10$^{-5}$ bar) maintained at room temperature. Background spectra were collected by measuring a mirror before the measurements.

## 1.4 INELASTIC NEUTRON SCATTERING MEASUREMENTS

Inelastic neuron scattering (INS) measurements were performed at the TOSCA spectrometer[1] at the ISIS Neutron and Muon Source, Rutherford Appleton Laboratory (Chilton, UK). High-resolution ($\Delta E/E$~1.25%) spectra, covering the broadband range (20-4000 cm$^{-1}$), were collected from the bulk polycrystalline sample (~1 g) at 10 K. The pulsed, polychromatic beam of neutrons collided with the sample and the scattered neutrons were Bragg reflected by a pyrolytic graphite analyser. Higher-order reflections beyond (002) were suppressed by a cooled ($T < 30$ K) Beryllium filter, acting as a longpass filter to analyse neutrons of a consistent final energy. Accordingly, neutrons with a final energy of ~32 cm$^{-1}$ were passed towards the detector array composed by thirteen $^3$He tubes with effective length of 250 mm. Five banks were located in forward direction (scattering angle ~45°) and five in backwards direction (~135°). The use of a low final energy translated into a direct relationship between energy transfer (ET, cm$^{-1}$) and momentum transfer ($Q$, Å$^{-1}$) such that ET ≈ $16Q^2$. Energy transfer and spectral intensity, i.e. $S(Q, \omega)$, were then obtained using the Mantid software.[2] The sample was wrapped in 4 cm × 4.6 cm aluminium sachet and placed into a 2.0 mm spaced flat aluminium cell, which was sealed with indium wire. To reduce the effect of the Debye-Waller factor on the experimental spectral intensity and allow comparison with the theoretical spectra, the sample cell was cooled to ~10 K by a closed cycle refrigerator. The INS spectrum was collected under vacuum over a duration of 5 hours.

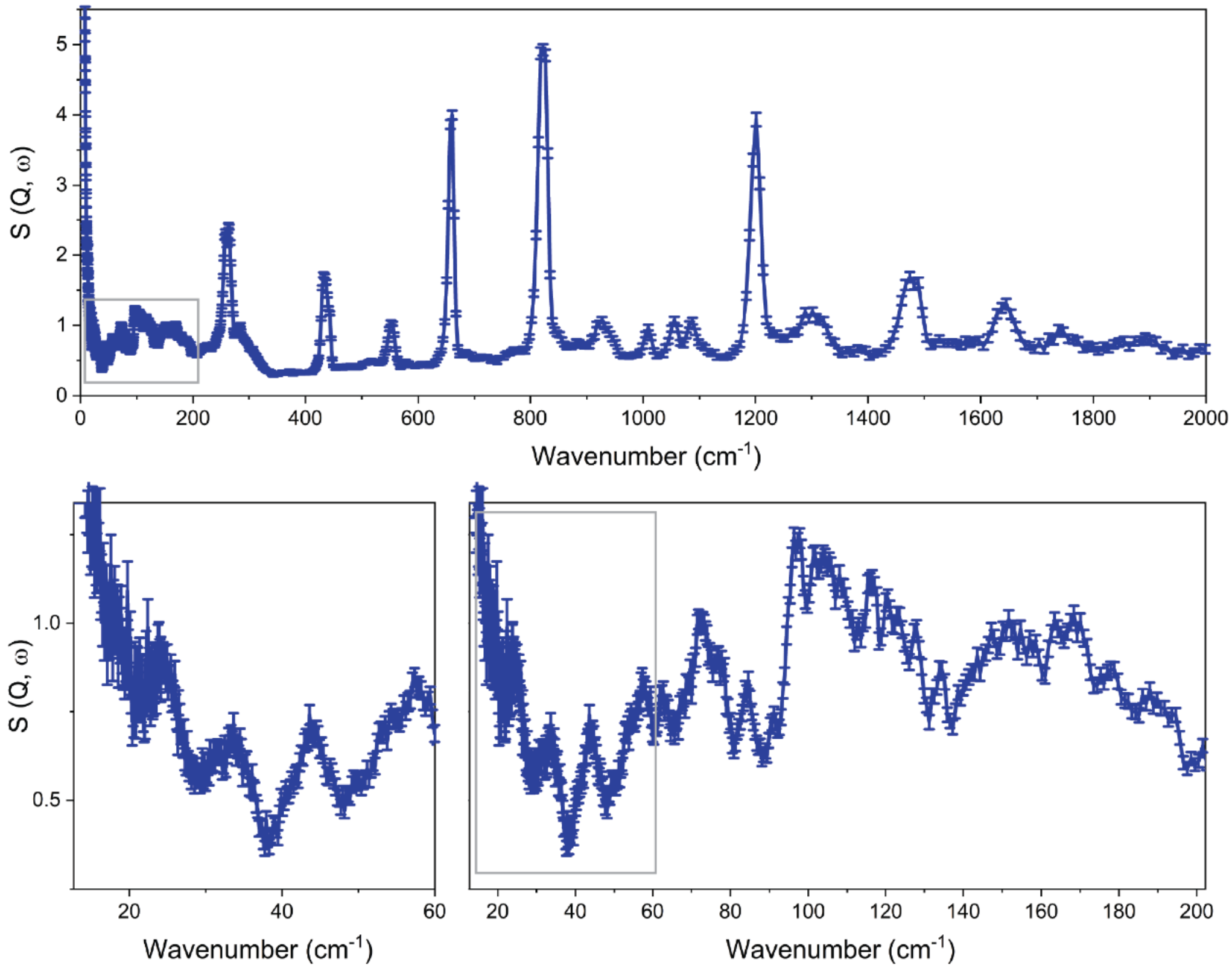

FIG S1: Inelastic neutron scattering (INS) data with error bars for a powder sample of ZIF-71.

The neutron guide upgrade of the TOSCA spectrometer, completed in 2017, has increased the neutron flux at the sample position by as much as 82 times. This upgrade improves the performance through faster measurements and by reducing the required sample mass.[3]

## 1.5 Density Functional Theory (DFT) Calculations

*Ab initio* density functional theory (DFT) calculations, in principle, evaluate the force-fields acting on the electrons. As the second derivative of the energy, the force constant considers the interactions emerging from electrons and nuclei within the molecule. In the DFT approach, the terms involving the Coulomb interactions between the electrons (exchange-correlation energy) are modelled by functionals employing different levels of theory. Here the crystalline orbital is approximated by a linear combination of Bloch functions, built from localized functions ("atomic orbitals"). These atomic orbitals are represented as linear combinations of Gaussian-type functions whose constant coefficients are defined by the input. Commonly used, for instance, is the B3LYP functional, a hybrid model proposed by Becke (B3)[4] to advance the gradient-corrected correlation of Lee et al. (LYP)[5]. Combined with the Grimme's dispersion correction (B3LYP-D3)[6], the DFT calculations yield theoretical IR spectra demonstrating the closest resemblance with experimental frequencies.[7] All electron basis sets were used for Zn, C, N, Cl, and H atoms, containing 12,480 (BS1) and 16,032 (BS2) local functions for the 816 atoms per unit cell.

First, the geometry optimization was carried out using a quasi-Newtonian algorithm and was considered complete when the calculation converged to the thresholds for both the RMS and maximum value for the force and atomic displacement, simultaneously. The corresponding thresholds were $3 \times 10^{-5}$ (RMS on gradient), $1.2 \times 10^{-4}$ (RMS on displacement), $4.5 \times 10^{-5}$ (largest component of gradient), and $1.8 \times 10^{-4}$ (absolute value of largest displacement). Subsequently, the IR frequencies were calculated at the special point $\Gamma = (0,0,0)$. Here, the dynamical matrix ("mass-weighted Hessian matrix") was obtained through numerically evaluating the first derivatives of the atomic gradients. The IR intensities were then calculated with the Berry Phase approach; for a detailed description of the performed calculation, we refer to the work of Pascale et al.[8] The continuous spectrum was obtained by fitting the calculated IR intensities with Lorentzian peak shapes with a FWHM of 10 $cm^{-1}$. To improve the match with experimental data, the calculated IR spectra were scaled by using a scaling factor of 0.98.[9] The INS spectrum was calculated from the output of the DFT frequency calculation using the Abins v1 execution in the Mantid software.[2]

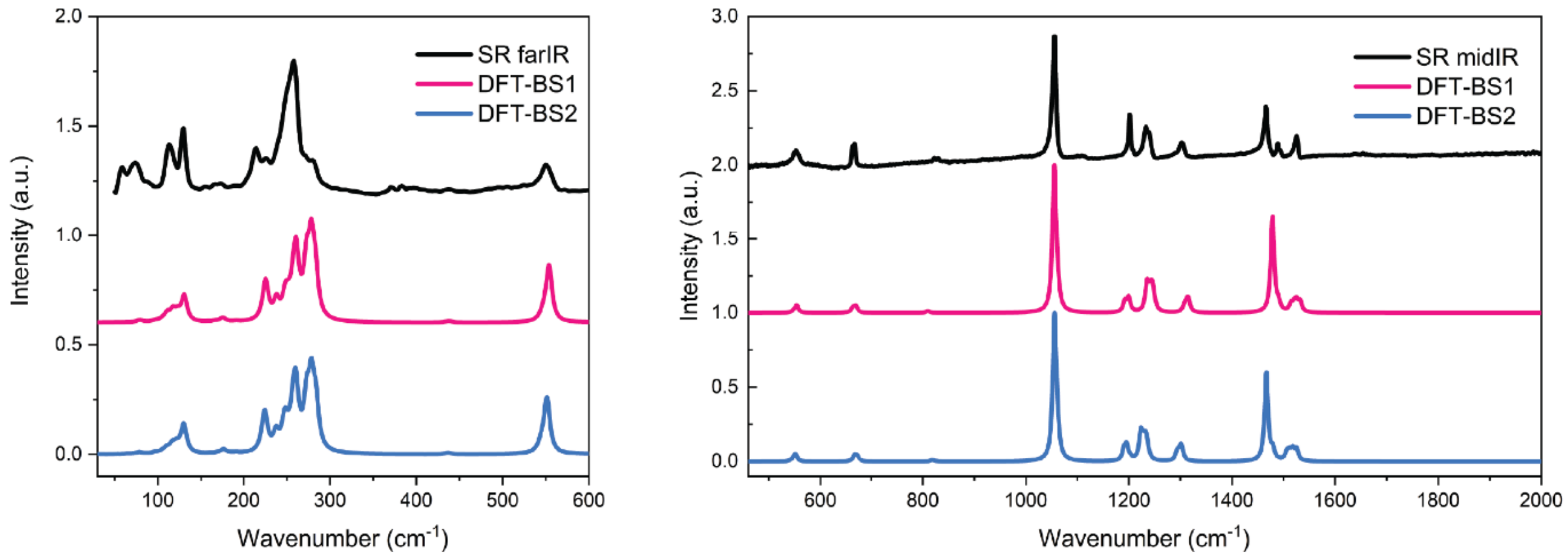

FIG S2: Comparison between infrared spectra of ZIF-71 obtained from synchrotron-radiation (SR) far-IR experiments and DFT simulations employing two different basis sets (BS1 and BS2).

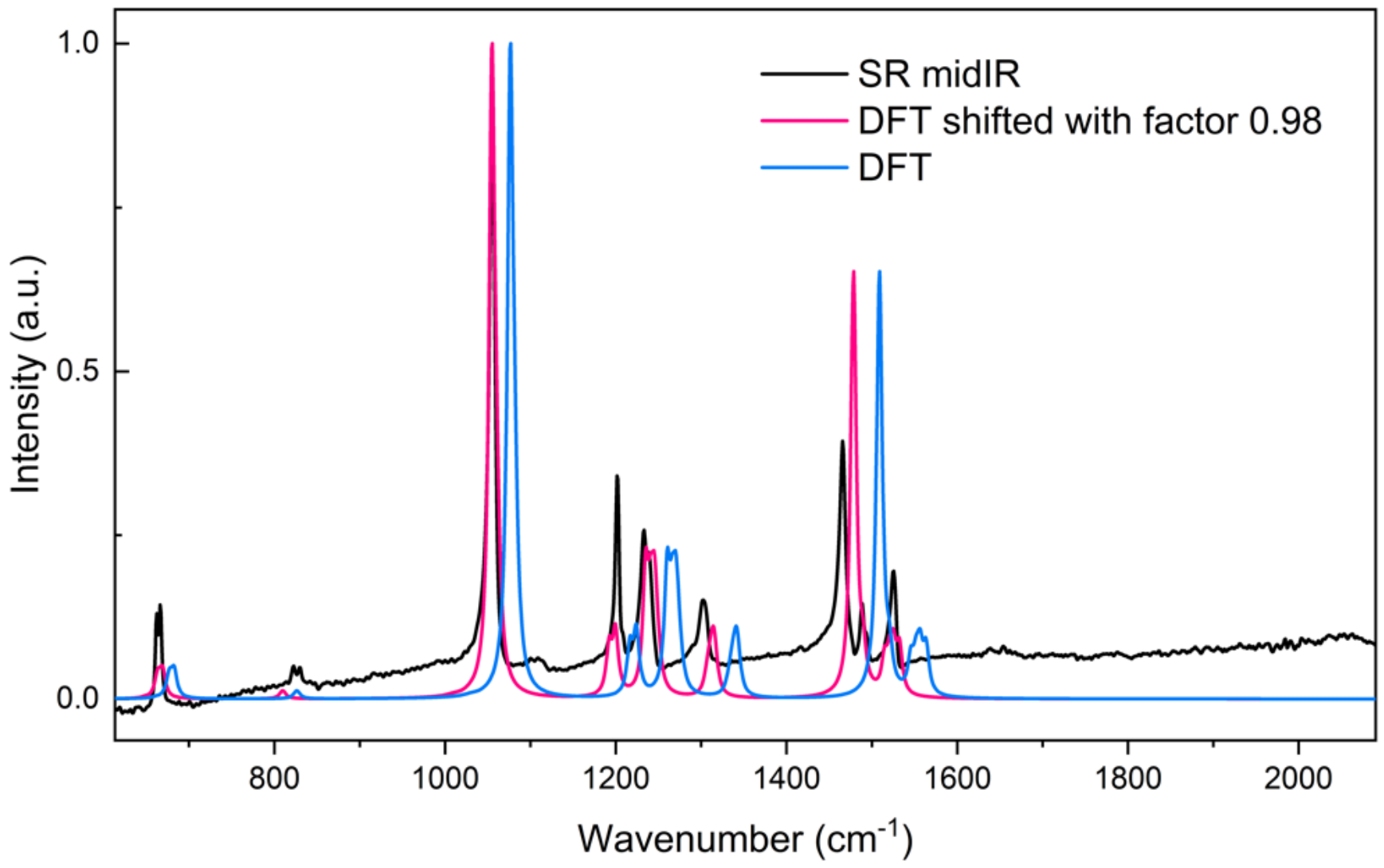

FIG S3: Bulk shift of DFT simulated spectrum versus the experimental mid-IR spectrum of ZIF-71.

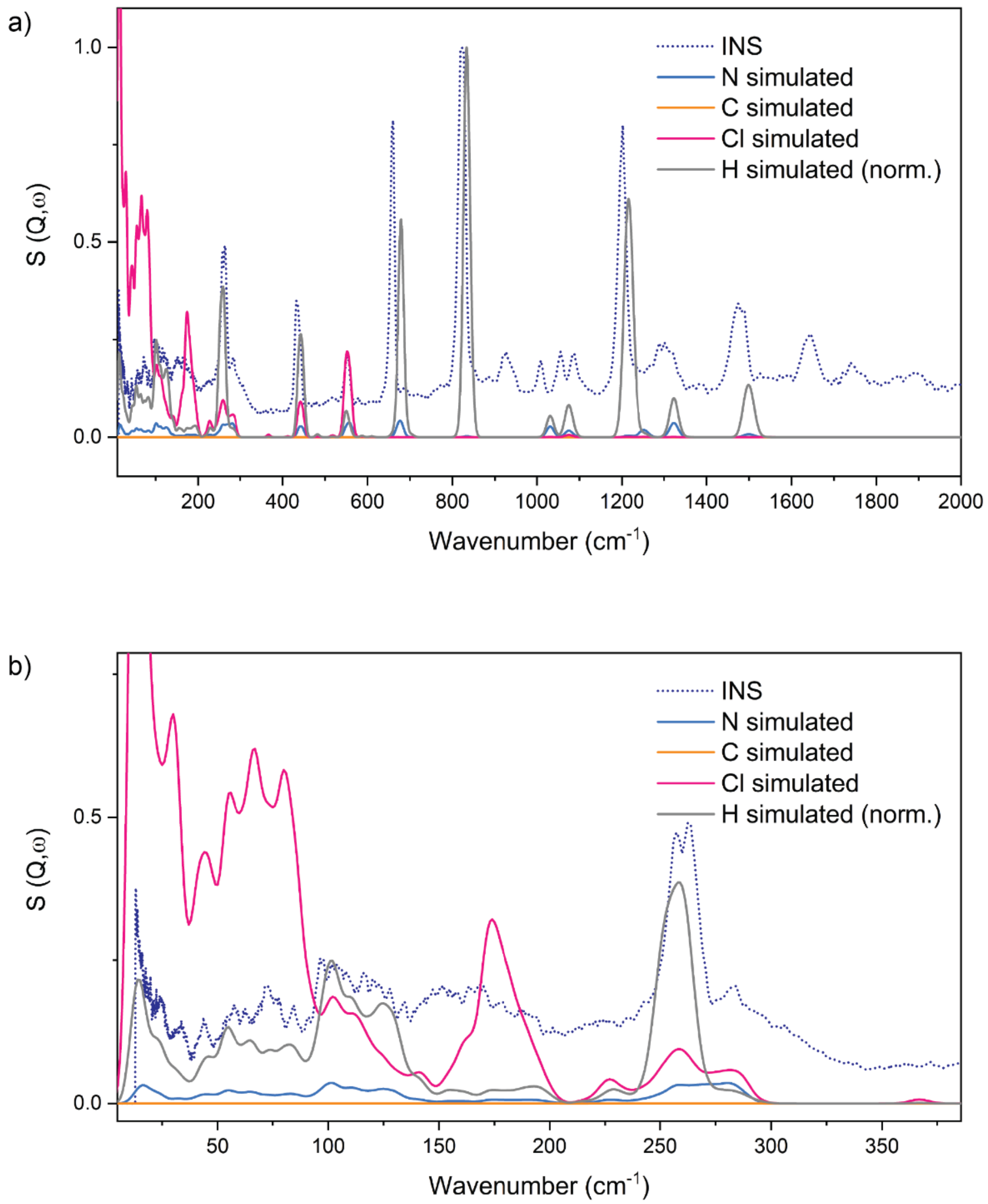

FIG S4: INS spectra derived from the phonon calculation compared with experimental measurement. No shift was applied, but the simulated spectrum for hydrogen (H) has been scaled down (normalization between 0 and 1) for better visualization.

## 1.6 NanoFTIR Measurements

The near-field optical measurements were performed with a neaSNOM instrument (neaspec GmbH) based on a tapping-mode atomic force microscopy (AFM) setup where the platinum-coated tip (NanoAndMore GmbH, cantilever resonance frequency 250 kHz, nominal tip radius ~20 nm) was illuminated by a broadband femtosecond laser. The coherent mid-infrared light was generated through the nonlinear difference-frequency combination of two beams from fiber lasers (TOPTICA Photonics Inc.) in a GaSe crystal. The spectra of two laser sources covering the range from 700 to 1400 $cm^{-1}$ and 1000 to 1600 $cm^{-1}$, respectively, were merged for the measurements. Demodulation of the optical signal at higher harmonics of the tip resonance frequency eliminated background contributions to yield the near-field signal, comprising amplitude and phase of the scattered wave from the tip. Employing a pseudo-heterodyne interferometric detection module, the complex optical response of the material is measured, where the real part refers to the nanoFTIR reflectance and the imaginary part depicts the nanoFTIR absorption spectrum. Each spectrum was acquired from an average of 10 Fourier-processed interferograms with 9 $cm^{-1}$ spectral resolution, 1024 points per interferogram, and 10-ms integration time per pixel. The sample spectrum was normalized to a reference spectrum measured on the silicon substrate. All measurements were carried out under ambient conditions.

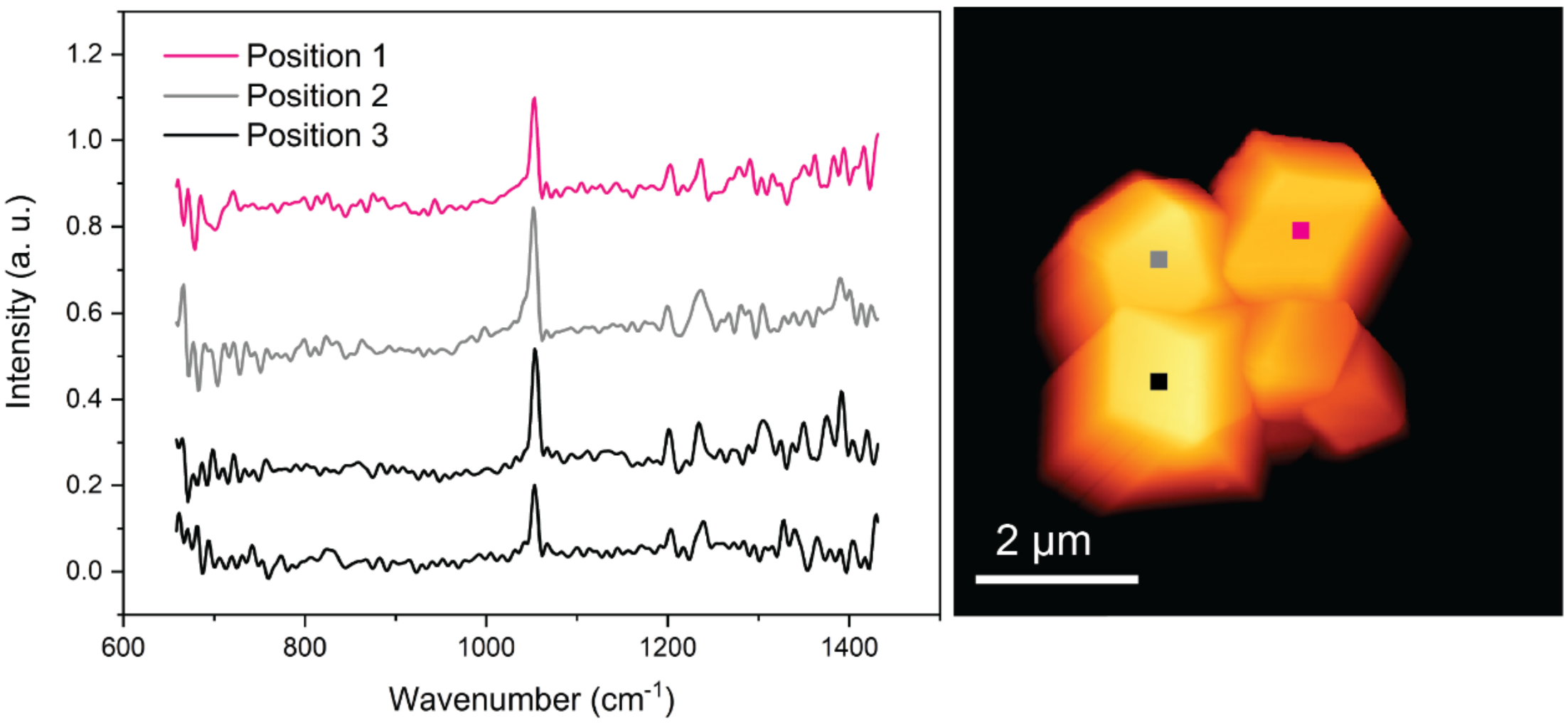

FIG S5: Local nanoFTIR spectra measured on individual ZIF-71 crystals reveal chemical homogeneity.

## 1.7 AFM Nanoindentation

AFM nanoindentation was performed with the Veeco Dimension 3100 instrument operating in indentation mode. A Bruker PDNISP probe with a cube-corner diamond indenter tip (cantilever spring constant 152 N/m, and contact sensitivity of 156.6 nm/V) was used. The nominal contact area was determined based on the indenter tip geometry established by Zeng and Tan.[10] The mechanical properties – Young's modulus (E) and hardness (H) – were derived from the set of indenter load-vs-displacement curves employing the Oliver and Pharr method, which is applicable to the cube-corner geometry.[11] For a detailed description of the AFM nanoindentation methodology for metal-organic framework crystals, we refer to the work of Zeng and Tan.[10]

## 1.8 Complete Assignment of Vibrational Modes of ZIF-71

Table S1: Complete assignment of all the vibrational modes of ZIF-71 between 0-170 $cm^{-1}$ (<5.1 THz), 171-600 $cm^{-1}$ (~5.1-18 THz), 601-1200 $cm^{-1}$ (~18-36 THz), and 1201-4000 (~36-120 THz) based on the DFT simulations. Abbreviations: MR membered-ring, ν stretching, $\nu_s$ symmetric stretching, $\nu_a$ asymmetric stretching, δ bending, $\delta_s$ scissoring, ρ rocking, τ twisting, ω wagging. Simulated vibrational modes that can be observed in experimental FTIR measurements are shown in grey.

Collective modes 0 - 170 cm$^{-1}$

| Frequency (cm$^{-1}$) | | | Intensity (a.u.) | | Assignment |
|---|---|---|---|---|---|
| FTIR | BS1 | BS2 | BS1 | BS2 | |
| | **9.46** | 11.61 | 1 | 1 | 8MR gate-opening |
| | **10.45** | 6.63 | 1 | 1 | Soft mode with shear deformation of 8MR and 6MR |
| | 10.52 | 12.7 | 8 | 6 | ρ opposite linkers in 4MR, asym gate-opening |
| | 17.25 | | 1 | | ρ opposite linkers in 6MR, asym gate-opening |
| 24 | **23.68** | 21.9 | 5 | 6 | 8MR gate-opening |
| | 24.21 | 23.94 | 2 | 1 | 8MR, 4MR shearing, ρ Cl |
| | 25.43 | 25.35 | 2 | 3 | 8MR breathing |
| 33 | 33.06 | 31.95 | 3 | 5 | $\nu_a$ Zn-N, N-C, ω Cl |
| | **37.42** | 40.18 | 2 | 1 | 6MR gate-opening |
| 43 | 40.81 | | 3 | | $\nu_s$ Cl-C-N, small 6MR breathing |
| | 46.11 | 47.47 | 2 | 1 | $\nu_s$ Zn-N, N-C, ω Cl, small 4MR and 8MR breathing |
| 53 | 51.21 | 53.74 | 5 | 4 | flapping of neighbouring linkers, asym. gate-opening, $\nu_a$ Zn-N |
| | 54.01 | 55.05 | 1 | 1 | $\nu_a$ Zn-N-C, 4MR breathing |
| | 55.68 | 58.95 | 3 | 1 | $\nu_a$ Zn-N, 4MR and 8MR shearing, ring breathing |
| | 61.39 | 62.87 | 2 | 2 | ω Cl, ring breathing, 4MR and 6MR shearing |
| 62 | 64.63 | 65.17 | 3 | 6 | ring breathing, ω Cl, deformation of 8MR and 4MR |
| | 66.35 | 68.83 | 1 | 3 | $\nu_a$ Zn-N, ring deformation, small pore deformation |
| | 77.37 | 75.83 | 1 | 5 | τ Cl-C-N, ring breathing, very small pore deformation |
| | 79.29 | 79.59 | 12 | 21 | τ Cl, 8MR aperture increase, $\nu_s$ Zn-N: 4MR pore stretching |
| 74 | 80.85 | 80.18 | 39 | 16 | $\nu_a$ Zn-N: 4MR pore deformation |
| | 86.14 | 83.13 | 5 | 3 | τ Cl, ν C-H, 8MR and 4MR pore deformation, 6MR breathing |
| | 91.56 | 85.82 | 2 | 1 | $\nu_a$ Cl-C-N, 6MR pore deformation, 8MR and 4MR deformation |
| | 93.31 | 90.73 | 1 | 1 | 4MR breathing, 6MR and 8MR small deformation |
| | 99.21 | 98.65 | 34 | 30 | ν ring, ω Cl, 4MR and 6MR aperture expansion |
| | 103.96 | 102.09 | 4 | 2 | 6MR breathing (small) |
| | 107.82 | 106.79 | 42 | 34 | 8MR breathing (small) |
| | 111.95 | 110.15 | 4 | 7 | 4MR shearing, 6MR deformation |
| 112 | 113.09 | 112.11 | 168 | 99 | 4MR and 6MR deformation, 8MR expansion |
| 112 | 119.66 | 118.4 | 218 | 162 | 4MR and 8MR deformation, 6MR expansion |
| | 121.91 | 121.28 | 19 | 19 | 8MR shearing, 6MR contraction |
| | 125.35 | 123.79 | 163 | 171 | 8MR breathing, 4MR and 6MR deformation |
| | 130.67 | 130.17 | 5 | 62 | $\delta_s$ Cl, ν C-N, 4MR and 8MR pore stretching |
| 129 | 133.15 | 132.41 | 600 | 651 | ν Zn-N, ν N-C-H, 4MR breathing |
| | 156.81 | 158.22 | 1 | 1 | δ Cl, τ Zn-N, 6MR pore deformation |
| | 161.39 | 163.13 | 10 | 10 | $\nu_a$ Zn-N, $\delta_s$ Cl, 6MR pore deformation |
| | 170.61 | 171.61 | 21 | 9 | $\nu_a$ Cl, $\nu_a$ ring |
| | 173.72 | 174.13 | 1 | 2 | $\nu_a$ ring, $\delta_s$ Cl |
| | 177.17 | 177.78 | 15 | 29 | ν Zn-N $\delta_s$ Cl, 4MR and 6MR pore deformation |

Vibrational modes 171 - 600 $cm^{-1}$

| Frequency ($cm^{-1}$) | | | Intensity (a.u.) | | Assignment |
|---|---|---|---|---|---|
| FTIR | BS1 | BS2 | BS1 | BS2 | |
| | 179.63 | 180.06 | 28 | 10 | δ Zn-N (tetrahedral deformation), ring deformation |
| 173 | 179.89 | 180.54 | 55 | 68.75 | δ Zn-N bending (tetrahedral deformation), ω Cl |
| | 182.72 | 183.88 | 2 | 1 | ν Zn-N, ring deformation |
| | 189.79 | 190.2 | 1 | 7 | δ Zn-N, ring stretching |
| 190 | 195.41 | 194.86 | 15 | 17 | δ Zn-N (tetrahedral deformation) |
| | 223.49 | 223.06 | 25 | 20 | ν Zn-N, 4MR breathing |
| | 229.01 | 227.38 | 169 | 272 | δ Zn-N, 8MR and 6MR deformation |
| 215 | 230.26 | 230.87 | 780 | 730 | δ Zn-N, pore deformation |
| | 236.01 | 234.78 | 2 | 2 | δ Zn-N, $\delta_s$ Cl |
| 225 | 242.95 | 241.98 | 422 | 402 | ν Zn-N |
| 247 | 253.13 | 251.87 | 434 | 597 | δ Zn-N, 4MR and 8MR expansion |
| | 254.65 | 253.26 | 77 | 71 | δ Zn-N, ω N-C-H |
| | 256.51 | 253.87 | 45 | 9 | δ Zn-N, ω C-H |
| | 256.79 | 254.08 | 5 | 5 | δ Zn-N, ω C-H |
| | 257.44 | 255.53 | 185 | 156 | δ Zn-N |
| | 262.35 | 260.99 | 138 | 188 | ω ring, 4MR asym. gate-opening |
| | 263.05 | 261.87 | 296 | 155 | $\nu_a$ Zn-N, ω ring |
| | 263.63 | 261.87 | 8 | 140 | $\delta_s$ Zn-N |
| 258 | 265.61 | 264.08 | 719 | 431 | ρ linker |
| 258 | 266.4 | 265.23 | 686 | 1082 | ω linker |
| | 269.2 | 268.46 | 87 | 90 | ν Zn-N, ring deformation |
| | 275.26 | 274.57 | 4 | 2 | $\nu_a$ Zn-N, ring stretch |
| 271 | 278.41 | 278.02 | 1165 | 1103 | $\nu_a$ Zn-N, ring and pore deformation |
| 281 | 284.04 | 283.7 | 1586 | 1504 | ρ Zn-N, ρ C-N-C, small 8MR breathing |
| | 288.77 | 288.94 | 969 | 1020 | τ N-C, small 6MR breathing |
| | 439.85 | 438.99 | 3 | 4 | ρ linker |
| | 444.45 | 443.66 | 7 | 6 | ρ linker |
| | 446.1 | 444.83 | 16 | 17 | $\nu_a$ N-C-Cl, only incomplete linkers |
| | 449.12 | 447.88 | 14 | 7 | $\delta_s$ linker |
| | 551.7 | 548.38 | 1 | 1 | $\delta_s$ N-C, 4MR deformation |
| | 552.08 | 548.59 | 2 | 10 | τ N-Zn-N, 4MR expansion |
| | 552.64 | 548.77 | 6 | 1 | τ N-Zn-N, |
| | 553.65 | 550.39 | 4 | 8 | ω N-Zn-N |
| | 554.53 | | 1 | | ω N-Zn-N, ring stretching |
| | 558.19 | 554.87 | 153 | 167 | ω C-N, τ C-N-Zn |
| | 560.49 | 557.14 | 14 | 15 | $\nu_s$ Zn-N (Zn fixed) |
| | 563 | 559.47 | 120 | 80 | $\nu_s$ Zn-N (Zn fixed) |
| | 563.53 | 559.85 | 41 | 95 | $\nu_s$ Zn-N (Zn fixed), ring shearing |
| | 564.74 | 560.98 | 173 | 216 | $\nu_s$ Zn-N (Zn fixed), ring deformation |
| 552 | 565.57 | 561.99 | 1044 | 1000 | $\nu_s$ Zn-N (Zn fixed), in-plane ring deformation |

Vibrational modes 601- 1200 $cm^{-1}$ (Ring deformations)

| Frequency ($cm^{-1}$) | | | Intensity (a.u.) | | Assignment |
|---|---|---|---|---|---|
| FTIR | BS1 | BS2 | BS1 | BS2 | |
| | 660.64 | 669.95 | 6 | 5 | out-of-plane ring deformation |
| | 661.19 | 670.36 | 19 | 1 | out-of-plane ring deformation |
| | 662 | 671.17 | 3 | 10 | out-of-plane ring deformation |
| | 663.36 | 672.4 | 5 | 10 | out-of-plane ring deformation |
| | 674.43 | 575.91 | 1 | 1 | out-of-plane ring deformation |
| | 675.12 | 677.05 | 99 | 8 | out-of-plane ring deformation |
| | 676.38 | 678.17 | 250 | 375 | out-of-plane ring deformation |
| 662 | 677.3 | 678.69 | 346 | 236 | out-of-plane ring deformation |
| | 677.77 | 678.96 | 157 | 354 | out-of-plane ring deformation |
| | 679.84 | 681.17 | 21 | 27 | out-of-plane ring deformation |
| | 680.36 | 681.96 | 65 | 56 | out-of-plane ring deformation |
| 666 | 682.82 | 684.58 | 972 | 761 | out-of-plane ring deformation |
| | 823.1 | 830.44 | 3 | 14 | out-of-plane ring deformation |
| | 824.39 | 831.07 | 58 | 1 | out-of-plane ring deformation |
| 821 | 825.09 | 832.29 | 70 | 116 | out-of-plane ring deformation |
| | 825.85 | 832.55 | 4 | 14 | out-of-plane ring deformation |
| 830 | 826.46 | 833.13 | 99 | 62 | out-of-plane ring deformation |
| | 827.29 | 835.16 | 29 | 81 | out-of-plane ring deformation |
| | 827.9 | 841.68 | 87 | 52 | out-of-plane ring deformation |
| | 1024.56 | 1027.14 | 12 | 8 | in-plane ring deformation |
| | 1028.59 | 1030.85 | 6 | 7 | in-plane ring deformation |
| | 1029.4 | 1031.71 | 4 | 2 | in-plane ring deformation |
| | 1029.54 | 1032.1 | 8 | 10 | in-plane ring deformation |
| | 1029.98 | 1032.58 | 4 | 5 | in-plane ring deformation |
| | 1033.14 | 1035.74 | 15 | 18 | in-plane ring deformation |
| | 1033.49 | 1063.7 | 3 | 8 | in-plane ring deformation |
| 1056 | **1076.38** | 1074.7 | 24028 | 24045 | in-plane ring deformation, ρ H-C-N |
| | 1080.26 | 1078.22 | 2317 | 2381 | in-plane ring deformation, ρ H-C-N |
| | 1081.76 | 1079.72 | 2493 | 2912 | in-plane ring deformation, ρ H-C-N |
| | 1084.43 | 1082.53 | 596 | 513 | in-plane ring deformation, ρ H-C-N |

Vibrational modes 1201 - 4000 $cm^{-1}$

| Frequency ($cm^{-1}$) | | | Intensity (a.u.) | | Assignment |
|---|---|---|---|---|---|
| FTIR | BS1 | BS2 | BS1 | BS2 | |
| 1201 | **1216.57** | 1212.3 | 1839 | 1699 | δ CH, $v_a$ Cl-C-N, edges of unit cell |
| | 1220.54 | 1216.47 | 1 | 1008 | δ CH, v C-N |
| 1201 | 1224.07 | 1218.27 | 2414 | 1666 | δ CH, v C-N |
| | 1229.18 | 1225.51 | 63 | 340 | δ CH, v C-N |
| 1233 | **1260.24** | 1245.5 | 3319 | 3431 | v C-N, ring breathing |
| | 1260.38 | 1245.82 | 1235 | 1174 | v C-N, ring breathing |
| | 1264.22 | 1249.53 | 46 | 184 | v C-N |
| | 1266.22 | 1251.13 | 2490 | 2191 | $v_a$ C-N |
| 1238 | 1270.56 | 1255.91 | 3175 | 4286 | $v_a$ C-N |
| | 1272.64 | 1237.63 | 548 | 411 | $v_a$ C-N |
| | 1273.35 | 1257.97 | 368 | 13 | v C-N |
| | 1276.42 | 1260..96 | 486 | 537 | v C-N |
| | **1332.01** | 1317.33 | 190 | 1367 | v C-N |
| | 1333.06 | 1320.68 | 11 | 275 | $v_s$ C-N |
| | 1334.13 | 1322.32 | 238 | 19 | v C-N |
| | 1334.48 | 1323.73 | 5 | 164 | $v_s$ C-N |
| 1302 | 1335.65 | 1324.99 | 440 | 1741 | v C-N |
| | 1336.15 | 1325.64 | 2 | 500 | $v_s$ C-N |
| | 1337.46 | 1325.64 | 78 | 5 | v C-N |
| | 1337.97 | 1326.37 | 300 | 83 | $v_s$ C-N |
| | 1339.14 | 1326.95 | 532 | 371 | $v_s$ C-N |
| | 1341.35 | 1328.42 | 1053 | 18 | $v_s$ C-N |
| | 1341.77 | 1328.6 | 346 | 17 | v C-N |
| 1465 | **1508.78** | 1493.54 | 16845 | 15950 | $δ_s$ N-CH |
| | 1519.82 | 1505.77 | 22 | 193 | $δ_s$ N-CH |
| 1488 | 1520.47 | 1506.45 | 1248 | 1509 | v C-N, δ CH |
| | 1543.99 | 1532.58 | 74 | 60 | $v_a$ C-C-N |
| 1518 | 1545.61 | 1534.23 | 1168 | 1307 | $v_a$ C-C-N, v C-C, |
| | 1551.32 | 1539.22 | 503 | 468 | $v_s$ N-C-N |
| | 1552 | 1539.91 | 630 | 844 | $v_s$ N-C-N |
| | 155.49 | 1545.15 | 302 | 316 | v C-C, v C-N, |
| 1525 | 1556.6 | 1546.23 | 1331 | 1399 | v C-C, v C-N, δ N-C-N |
| | 1562.84 | 1552379 | 488 | 519 | $v_s$ N-C-Cl (edge of unit cell) |
| | 1563.73 | 1553.68 | 1273 | 1423 | $v_s$ N-C-Cl (edge of unit cell) |
| | 3277 | 3251.24 | 158 | 134 | ρ CH |
| | 3277.05 | 3251.28 | 15 | 10 | v CH |
| | 3283.44 | 3260.16 | 31 | 38 | v CH |
| | 3283.78 | 3260.53 | 12 | 14 | v CH |
| | 3296.46 | 3270.69 | 19 | 31 | v CH |
| | 3296.78 | 3270.95 | 2 | 3 | v CH |
| | 3305.46 | 3281.78 | 74 | 86 | v CH |
| | 3305.78 | 3281.96 | 4 | 3 | v CH |